# 29P/Schwassmann-Wachmann 1: A Rosetta Stone for Amorphous Water Ice and CO <-> CO₂ Conversion in Centaurs and Comets?

C.M. Lisse[1], J.K. Steckloff [2,3], D. Prialnik[4], M. Womack[5,6], O. Harrington Pinto[5], G. Sarid[7], Y.R. Fernandez[5,8], C.A. Schambeau[5,8], T. Kareta[9], N.H. Samarasinha[2], W. Harris[10], K. Volk[10], L.M. Woodney[11], D.P. Cruikshank[5], S.A. Sandford[12]

[1]Planetary Exploration Group, Space Department, Johns Hopkins University Applied Physics Laboratory, 11100 Johns Hopkins Rd, Laurel, MD 20723  carey.lisse@jhuapl.edu

[2]Planetary Science Institute, 1700 E Fort Lowell Road, Suite 106, Tucson, AZ 85719 jordan@psi.edu, nalin@psi.edu

[3]Department of Aerospace Engineering and Engineering Mechanics, University of Texas at Austin, Austin, TX, 78712-1221 steckloff@utexas.edu

[4]Department of Geosciences, Faculty of Exact Sciences, Tel Aviv University, Tel Aviv Israel  dinak@tauex.tau.ac.il

[5]Department of Physics, University of Central Florida, Orlando, FL 32816  mariawomack@gmail.com, yan@physics.ucf.edu, oharrington@knights.ucf.edu, dpcruikshank@comcast.net

[6]National Science Foundation, 2415 Eisenhower Avenue, Alexandria, Virginia 22314, USA

[7]SETI Institute, 339 Bernardo Ave, Suite 200, Mountain View, CA 94043 galahead@gmail.com

[8]Florida Space Institute, University of Central Florida, Orlando, FL 32816  Charles.Schambeau@ucf.edu

[9]Lowell Observatory, 1400 West Mars Hill Road, Flagstaff, AZ 86001, USA tkareta@lowell.edu

[10]Lunar & Planetary Laboratory, University of Arizona, Tucson, AZ  wharris@lpl.arizona.edu, kvolk@lpl.arizona.edu

[11]Department of Physics, Cal State University San Bernardino, 5500 University Parkway, San Bernardino, CA 92407 woodney@csusb.edu

[12]Astrophysics Branch, Space Sciences Division, NASA/Ames Research Center, Moffett Field, CA, 94035, USA scott.sandford@nasa.gov

28 Pages, 3 Figures, 2 Tables

# Abstract

Centaur 29P/Schwassmann-Wachmann 1 (SW1) is a highly active object orbiting in the transitional "Gateway" region between the Centaur and Jupiter Family Comet regions. SW1 is unique among the Centaurs in that it experiences quasi-regular major outbursts and produces CO emission continuously; however, the source of the CO is unclear. We argue that due to its very large size (~32 km radius), SW1 is likely still responding, via amorphous water ice (AWI) conversion to crystalline water ice (CWI), to the "sudden" change in its external thermal environment produced by its dynamical migration from the Kuiper belt to the Gateway Region at the inner edge of the Centaur region at 6 au. It is this conversion process that is the source of the abundant CO and dust released from the object during its quiescent and outburst phases. If correct, these arguments have a number of important predictions testable via remote sensing and *in situ* spacecraft characterization, including: the quick release on Myr timescales of CO from AWI conversion for any few km-scale scattered disk KBO transiting into the inner system; that to date SW1 has only converted between 50 to 65% of its nuclear AWI to CWI; that volume changes upon AWI conversion could have caused subsidence and cave-ins, but not significant mass wasting or crater loss; that SW1's coma should contain abundant amounts of CWI $CO_2$-rich "dust" particles; and that when SW1 transits into the inner system within the next 10,000 years, it will be a very different kind of JFC comet.

**1. Argument.** Centaur 29P/Schwassmann-Wachmann 1 is a relatively large (~32 km radius, Schambeau *et al.* 2021a, Bockelee-Morvan *et al.* 2022[1]) icy planetesimal residing in a nearly circular orbit just beyond the orbit of Jupiter. It is well known for its unusually high level of CO production and dust emission activity (but not conspicuously for any significant $H_2O$ emission activity) and frequent outbursts (Senay & Jewitt 1994; Gunnarsson *et al.* 2008, Trigo-Rodriguez 2008, 2010; Hosek *et al.* 2013; Wierzchos & Womack 2020; etc.). Dynamically, Centaurs are an unstable transitional population and they represent the middle state between the long-lived reservoir of icy Kuiper Belt Objects (KBOs) in the outer solar system and the quickly evolving Short Period comet (SP) population in the inner solar system (see, e.g., reviews by Dones *et al.* 2015, Peixinho *et al.* 2020, Fraser *et al.* 2022). SW1 currently resides in a 'Gateway' orbit: a collection of dynamical orbits that facilitate dynamical migration between these two populations (Sarid *et al.* 2019; Steckloff *et al.* 2020; Seligman *et al.* 2021). Sarid *et al.* 2019 showed that SW1's low-eccentricity orbit just exterior to Jupiter is typical for Centaurs transitioning to Jupiter family comet (JFC) orbits and is very likely to undergo this transition within the next ~10 Kyr.

Starting in the 1970s, astronomical infrared spectral studies detected absorption features indicative of ices containing $H_2O$, CO, $CO_2$, $CH_4$, $H_2CO$, $NH_3$, and $CH_3OH$ (Schwartz *et al.* 1973, Merill *et al.* 1976, Soifer *et al.* 1979, Allamandola *et al.* 1992, Lacy *et al.* 1998) while investigating the composition of icy molecular cloud cores, the precursors to solar systems and their icy planetesimals. The spectral and physical properties of these ices, including their sublimation and condensation behaviors, were subsequently studied in the laboratory (c.f. Lisse *et al.* 2021 & references therein) to investigate the possible makeup of these clouds, and the plausibility that the laboratory ice analogues could be present in the molecular cloud at the ambient temperatures estimated from their spectroscopy. This work has been bolstered by spectral studies of ices detected on planetary satellites, Centaurs, Pluto, and KBOs in the outer Solar System like $H_2O$, $CH_4$, $N_2$, CO, $CO_2$, $CH_3OH$, HCN, $NH_3 \cdot nH_2O$, and $C_2H_6$ (Cruikshank *et al.* 1998; Grundy *et al.* 2006; Brown *et al.* 2007; Barucci *et al.* 2008, 2011) and those detected in cometary comae as the products of sublimative mass loss from the parent nucleus, such as $C_2H_2$, $C_3H_8$, $SO_2$, and $O_2$ (Bieler

---

[1] For an assumed $p_v = 0.04$. There is a smaller published value of $R_{nuc} = 20 \pm 4$ km, $p_v = 0.13 \pm 0.04$ obtained by Cruikshank & Brown 1983 during a period of low SW1 nuclear activity. Re-analyzing the fluxes presented in the 1983 paper with modern thermophysical models, we find $R_{nuc} = 50 \pm 8$ km, $p_v = 0.02 \pm 0.01$. Thus throughout this paper when we state $R_{nuc} = 32$ km, we are really saying $R_{nuc} = 32 ^{-14/+28}$ km. The net result is to broaden the estimated AWI conversion timescales to 30 - 200 Myr (from 60 - 100 Myr), still very much comfortably >> the 10 Myr dynamical JSUN region crossing time (and a few Myr JS-region crossing time) for all currently plausible values of $R_{nuc}$.

*et al.* 2015, Mall *et al.* 2016). What is of most importance for this study of SW1 are the 3 most common species, $H_2O$, $CO$, and $CO_2$, which can be present as either pure or mixed ices, and in non-crystalline, amorphous, low-temperature kinetic product form (e.g., amorphous solid water ice = AWI; Stevenson *et al.* 1999; Kimmel *et al.* 2001b; Dohnálek *et al.* 2003; Raut *et al.* 2007a,b) or in higher temperature, crystalline, lowest thermodynamic energy state form (e.g. crystalline water ice = CWI). For SW1, the properties and behavior of AWI and CWI are highly relevant, as there are a series of phase changes at low pressure in the 80 – 130 K temperature region from AWI to CWI, and $T_{LTE}$[2] for SW1 where it currently resides at 6 au is ~115 K (after spending ~4.5 Gyr in the Edgeworth Kuiper Belt (EKB) at $T_{LTE}$ = 30 – 40 K). The crystallization transformation of *pure* AWI is moderately exothermic, but common volatile impurities in comets could render the process neutral to moderately endothermic (Kouchi & Sirono 2001). By contrast, CO ice sublimates fully in the 20 – 30 K range (Davidsson *et al.* 2021, Lisse *et al.* 2021, 2022; Steckloff *et al.* 2021a) and $CO_2$ transforms from its amorphous to crystalline phase by ~30K (Escribanoa *et al.* 2013). We therefore expect SW1 to mainly consist of some sort of mixture of AWI, CWI, and crystalline $CO_2$ ices intermixed with rocky material and the minor ice species.

SW1 is special because it is very large compared to the typical "Gateway" Centaur/Comet. Sarid *et al.* (2019) estimate that only ~4% of the objects reaching the Gateway are this size or larger. Its volume reserves of AWI (~$4/3\pi$ x [32 km]$^3$ ) are thus large enough versus the input solar energy flux (= $\pi$ x [32 km]$^2$ x [1- $A_{bond}$$^3$]), that $\tau_{thermal}$, the time it takes to convert all its AWI -> CWI (the proposed activity driving process for Centaurs occurring inside 10 au; Prialnik *et al.* 1995, Jewitt 2009, Li *et al.* 2020) is 60 - 100 Myr (see Section 2.1). This is much longer than the ~10 Myr it typically takes a KBO to travel from the outer solar system to 6 au (Volk & Malhotra 2008, Prialnik & Rosenberg 2009, Sarid *et al.* 2019, Di Sisto & Rossignoli 2020, Gkotsinas *et al.* 2022) and the few Myr the KBO resides inside 10 au (Saturn's orbit), meaning that it has not yet exhausted the supply of any AWI it may have had while residing in the Kuiper Belt region, and it could ***still*** be undergoing AWI -> CWI conversion today.

---

[2] $T_{LTE}$ = 282/sqrt($r_h$), the equilibrium temperature achieved by a uniformly illuminated blackbody at distance $r_h$ from the Sun.
[3] $A_{bond}$ = $L_{scattered}$/($L_{scattered}$ + $L_{emitted}$), where L = luminosity, or the sum total energy output, across all wave-lengths and angles, of a body. $A_{bond}$ varies from 0.01 at 0.5 um to 0.2 at 10 um (Schambeau *et al.* 2015, 2021a), but is always << 1 and thus scattering of incident sunlight is relatively unimportant for SW1's overall energy balance.





By contrast, any AWI in four other, much smaller Centaurs ($R_{nuc}$ = 1 - 6 km) in similar Gateway orbits around the Sun (P/2010 TO20 LINEAR-Grauer, 423P/Lemmon, 2016 LN8, and 2019 LD2 (ATLAS); Sarid *et al.* 2019; Steckloff *et al.* 2020, Bolin *et al.* 2021, Kareta *et al.* 2021, Schambeau *et al.* 2021b) would have crystallized long ago, within 0.1 - 6 Myr of the start of their dynamical migration from the Kuiper Belt to the dynamical Gateway near Jupiter's orbit. Thus these other objects, like all known km-sized JFCs, should be depleted in AWI, and their current activity is likely dominated by the sublimation behavior of crystalline water ice and its entrained impurities.

In the Sections 2 - 7, we outline the timescale calculations and supporting arguments for the AWI conversion hypothesis. In Section 8, we discuss how comparative remote sensing studies of the activity patterns of Centaurs and Gateway objects versus size, heliocentric distance, and dynamical age could shed light on whether or not their mass loss is driven by thermal wave interior propagation and AWI conversion. In Section 9, we examine the processes and morphologies created by AWI conversion that an *in situ* spacecraft mission could uniquely search for.

## 2.    Supporting Arguments.

### 2.1    Thermal Timescales.
Central to our arguments is an understanding of the thermal history of heat flow through an icy, undifferentiated, yet geologically complex object like SW1. The expected dependence of the conversion timescale on body size can be understood in a number of ways: we start off with a very simple back of the envelope energy balance argument; then progress to a moderately simple argument invoking the results expected for heat flow in a uniform body of finite thermal diffusivity; and finally graduate to the much more sophisticated modeling of Prialnik *et al.* 2004, 2008, and Prialnik & Rosenberg 2009, which includes energy balance, finite heat flow, layering, and sublimative effects. We present all of these because the first two simplified approaches add value for the reader to understand the physics of the problem.

We first establish the thermal time constant for a response to a sudden (< ~10 Myr) change in the outside temperature/local insolation environment of SW1 **using simple energy balance**. Implicit in this argument is that the time-limiting step is the delivery of energy via insolation to the body, not the flow of heat from the surface to the interior of the body, and that energy inputs from short-lived radionuclides are negligible (Prialnik 2021, Steckloff *et al.* 2021a). We argue via analogy





using the modeling solutions found for a body similar to SW1 when it was in the Kuiper Belt: Arrokoth, a body about the same size as SW1, and the subject of the first-ever close flyby of a cold-classical KBO by a spacecraft on 01 Jan 2019. The New Horizons mission flew within 3500 km of the object and conducted many spectrophotometric imaging studies, finding a highly flattened, inactive object with about 1/2 the width and 1/3 the thickness of a sphere that has remained at ~45 au from the Sun since its formation as a contact binary some 4.5 Gyr ago (Stern *et al.* 2019, McKinnon *et al.* 2020). Arrokoth has been the subject of several studies of its thermal behavior since (Davidsson 2021, Lisse *et al.* 2021, 2022, Prialnik 2021, Steckloff *et al.* 2021a), in order to try and understand why there was no surrounding gas coma produced by sublimative activity of easily vaporized hypervolatiles like nearly pure phases of $N_2$, $CH_4$, or CO (Gladstone *et al.* 2022, Lisse *et al.* 2022). The results of these studies found that it took ~20 Myr once the protoplanetary disk (PPD) cleared for Arrokoth to lose all its hypervolatiles.

Knowing Arrokoth's thermal timescale for hypervolatile loss, we can scale it to determine SW1's thermal timescale for hypervolatile loss. In general, the timescale for which an ice inside a KBO, Centaur, or comet is transformed solely via radiative solar heating should go as

$\tau_{thermal}$ ~ [Total Reservoir of ices to be transformed/Rate of Energy Input for Transformation]

$\quad$ ~ $V_{nuc}$/ $A_{nuc}$ [Surface Area for absorbing solar radiation]

$\quad$ = $4/3\pi R_{nuc}^3/\pi R_{nuc}^2$ ~ $R_{nuc}$ [for a spherical body[4]]

Using Arrokoth's measured dimensions from the New Horizons flyby (Stern *et al.* 2019) and Prialnik 2021's ~20 Myr timescale for loss of hypervolatiles we can now produce a timescale estimate for SW1's loss of hypervolatiles. A spherical 32 km radius SW1-sized object should lose its hypervolatiles in the Kuiper Belt due to the same sudden change in the local equilibrium temperature $T_{LTE}$ (as the PPD cleared $T_{LTE}$ increased from ~20 to 40 K) in

$\tau_{thermal,SW1}$ ~ $\tau_{thermal,Arrokoth}$ × $(V_{nuc,SW1}/V_{nuc,Arrokoth})$ / $(A_{nuc,SW1}/A_{nuc,Arrokoth})$

---

[4] But Arrokoth is a very NON-spherical body (Stern *et al.* 2019, Keane *et al.* 2022), so we use its actual dimensions in our calculation here.





$$\sim 20 Myr \; x \; \frac{\left[\frac{4/3 \pi (32\,km)^3}{35km \; x \; 17km \; x \; 12km}\right]}{\left[\pi \; x \; 32km^2 \big/ (35km \; x \; 17\,km \; x \; 0.5)\right]} = \; 36 \; Myr$$

where the factor of 0.5 has been introduced into the term for Arrokoth's surface area to allow for the fact that it is only face-on to the Sun for about 1/2 of its orbit. Thus, from exposed surface to volume energy balance considerations alone, SW1 should respond to a sudden large change in incoming insolation energy about twice as slowly as Arrokoth does, mostly as a result of SW1 being about 2 times larger in effective radius than Arrokoth. By contrast, this calculation suggests that a 1 km radius spherical "typical" comet nucleus will respond to a sudden insolation change by processing any easily vaporized ices within $\sim 1$ Myr, while even a large Halley-sized $\sim 6$ km radius body will do so within $\sim 6$ Myr.

If instead we consider that thermal heat transport into and out of SW1 is the rate limiting step controlling AWI to CWI conversion, we find that the timescale required to convert SW1's will trend as $R_{nuc}^2$. This is because the heat-depth penetration distance ($l_{heat}$) solution for the 1-D radial heat flow equation assuming a step-function heat input is given by

$$l_{heat} = \sqrt{\mathcal{H} t_{warm}}$$
$$l_{heat}^2 = \mathcal{H} t_{warm}$$

where $\mathcal{H}$ is the thermal diffusivity of the material (typically $\sim 1$ x $10^{-7}$ m$^2$/s for cometary materials; Prialnik *et al.* 2004, Steckloff *et al.* 2021b) and $t_{warm}$ is the elapsed time. For this case, we consider the time SW1 has spent close enough to the sun for the surface to be warm enough for AWI to crystallize ($t_{warm}$), and compare to the radius of the nucleus, $R_{nuc}$ ; so long as the heat penetration depth is less than the radius of the object then it is possible for AWI to survive in the object's interior. Thus, AWI can survive so long as

$$\frac{R_{nuc}^2}{\mathcal{H}} > t_{warm}$$

From direct calculation using $\mathcal{H}$ = 1 x $10^{-7}$ m$^2$/s, AWI should be able to survive on the order of 290 Myr in the interior of SW1 while resident in the Gateway. Alternatively, Prialnik (2021) found that the crystallization timescale for Arrokoth, with $R_{nuc\_eff} = (35 \times 17 \times 12)^{1/3} = 19$ km is ~22 Myr; scaling this result to the ~32 km radius size of SW1, for a $t_{warm}$ that scales as radius squared, one finds a timescale of ~62 Myr. Both of these estimates are in stark contrast to the timescales for more typical comet nuclei; using $R_{nuc}^2$ scaling, an ~1 km radius nucleus would crystallize all its AWI on timescales of ~0.06 Myr and an ~6 km radius Halley-sized nucleus would exhaust its AWI after ~2 Myr of continuously residing at ~6 au.

Finally, we present the sophisticated modeling treatment of Prialnik 2021, which follows the changing internal structure of an icy SW1 nucleus reacting to solar heating, starting with a composition of CO-laden amorphous ice and rock, until the ice crystallizes throughout the body. The code solves coupled differential equations for energy and mass flows simultaneously, while allowing for internal heat sources like the heat of phase change and bodies consisting of many different ice species, each with their own effective heat of sublimation and crystallization. Energy can diffuse into the interior via solid state conduction, radiation, and/or gas advection. We do not go into the details of the calculations more here, but refer the reader to the most recent reviews of the models found in Prialnik *et al.* 2004, Prialnik *et al.* 2008, and Prialnik & Rosenberg 2009.

### Table 1 – Prialnik Model Parameters

| Parameter | Value |
|---|---|
| Ice heat capacity | $7.5 \times 10^4$ T + $9.0 \times 10^5$ erg g$^{-1}$ K$^{-1}$ |
| Dust heat capacity | $1.3 \times 10^7$ erg g$^{-1}$ K$^{-1}$ |
| Amorphous ice thermal conductivity | $2.35 \times 10^2$T + $2.82 \times 10^3$ erg cm$^{-1}$ s$^{-1}$ K$^{-1}$ |
| Crystalline ice thermal conductivity | $5.67 \times 10^7$/T erg cm$^{-1}$ s$^{-1}$ K$^{-1}$ |
| Dust thermal conductivity | $2 \times 10^4$ erg cm$^{-1}$ s$^{-1}$ K$^{-1}$ |
| Crystallization rate | $1.05 \times 10^{13}$ e$^{-5370/T}$ s$^{-1}$ |
| Latent heat of ice sublimation | $2.8 \times 10^{10}$ erg g$^{-1}$ |
| Dust specific density | 3.25 g cm$^{-3}$ |
| Average pore size | 0.1 cm |

N.B: The thermal conductivity is corrected for porosity ($\psi$) by a factor ($1-\psi^{2/3}$) and includes radiative conductivity in pores.

In applying the sophisticated Prialnik 1-D code with the parameters listed in Table 1 to the case of SW1, we considered models of different radii, adopting SW1's orbit and an albedo of 0.062, and obtained upper limits for the time required for full crystallization by assuming that no heat is released by the crystallization process. Somewhat shorter time scales were obtained by assuming AWI crystallization to release heat in the amount of -45 kJ/kg. The results of this numerical model, assuming a chondritic abundance SW1 containing a 1:1 mixture of ice to rock with mean density

= 0.5 g/cm³, corresponding to a porosity of 0.65, are shown in Figures 1 & 2 and Table 2. The rate of advance of the crystallization front for a 30 km radius object may be inferred from the left panel of Fig.1, which shows the residual AWI volume as function of time. The right panel shows the time for total conversion of all AWI in a body of radius $R_{nuc}$.

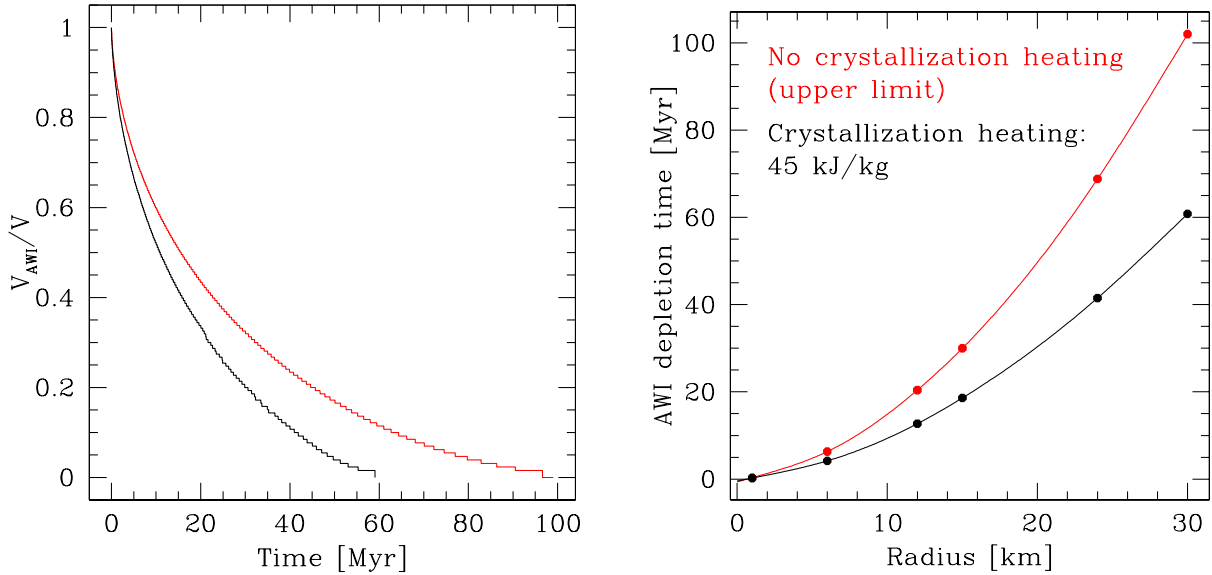

**Figure 1 – Prialnik model results for AWI depletion** (via conversion to CWI) for SW1-like bodies of bulk chondritic abundance and 1:1 ice:rock ratio. In both plots the two curves are the results for assuming two different enthalpies for the AWI -> CWI transition: **red** for $\Delta H_{AWI\ conversion} = 0$ energy released, which produces the slowest transformation rate, as there is no additional energy boost from the phase change itself to continue the process; **black** for $\Delta H_{AWI\ conversion} = -45$ kJ of heat released per kg of CWI produced. (*Left*) Fraction of AWI left after a given time for a spherical 32 km radius SW1-like body. Note that after 10 Myr, only ~35% of the AWI in the $\Delta H = 0$ kJ/kg and 50% of the AWI in the $\Delta H = -45$ kJ/kg released case has transformed. (*Right*) Time for depletion of all AWI throughout the entire body. The $\Delta H = 0$ curve follows an approximate $R_{nuc}^{1.65}$ law & the $\Delta H = -45$ kJ/kg curve an approximate $R_{nuc}^{1.61}$ law.

### Table 2 – Prialnik Model: Time in Myrs Required to Convert All AWI to CWI

| Radius (km) | $\tau_{AWI->CWI}$ (Myr) for $\Delta H_{AWI->CWI} = 10^{-3}$ kJ/kg ~ 0 | $\tau_{AWI->CWI}$ (Myr) for $\Delta H_{AWI->CWI} = -45$ kJ/kg |
|---|---|---|
| 1 | 0.33 | 0.24 |
| 6 | 6.3 | 4.2 |
| 12 | 20 | 13 |
| 15 | 30 | 19 |
| 24 | 69 | 42 |
| 30 | 100 | 61 |

The model time estimates for SW1 to lose all its AWI fall between 60 and 100 Myr (Fig. 1a), while a "typical" 1 km radius sized Gateway Centaur would lose all its AWI in the first 0.2 – 0.3 Myr, and even a 6 km radius "Halley-sized" Gateway Centaur would convert all its AWI in 4 – 6 Myrs (Fig. 1b).

It is noteworthy that for the adopted composition, $CO/H_2O$ = 0.05 by mass, *the estimated current production rate of CO for SW1, ~4e28 mol/sec, released from the AWI exceeds the sublimation rate of water ice at the surface by many orders of magnitude* (Figure 2) and requires that the AWI conversion process be mildly exothermic, $\Delta H_{AWI\ conversion}$ ~ -45 kJ/kg (or ½ the maximal $\Delta H_{AWI\ conversion}$ = 90 kg/kg for pure water ice, Jewitt 2009). The estimated CO and $H_2O$ gas production rates become comparable only when the AWI crystallization front has receded to ~0.6 of the radius (after 30 Myr residence time at 6 au, $\Delta H_{AWI\ conversion}$ = -45 kJ/kg; after 60 Myr residence time at 6 au, $\Delta H_{AWI\ conversion}$ = 0).

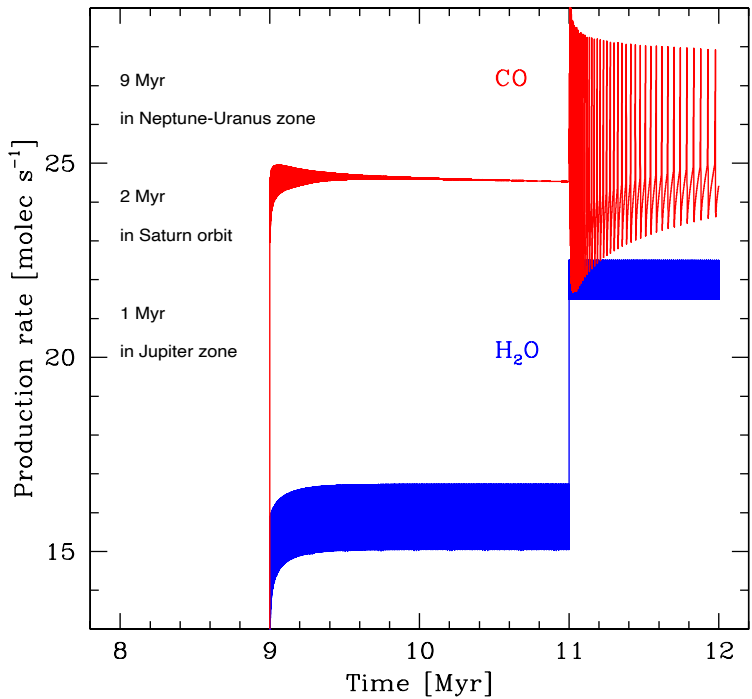

**Figure 2.** - Relative CO and $H_2O$ gas production rates for the Prialnik model of Figure 1, assuming a 32 km radius, 5% $CO/H_2O$, 1:1 Ice/Rock (by mass) body that has spent 4.56 Gyr in the Kuiper Belt, then 9 Myr in the Neptune- Uranus region, 2 Myr near Saturn, and finally the last 1 Myr near Jupiter at 6 au. Other dynamical trajectories are possible, for example a body that has slowly but monotonically inspiralled from ~40 au to ~6 au rather than moved stochastically, and they produce similar $Q_{CO}/Q_{H_2O}$ > $10^3$ production ratios. Exploring the expected relative outgassing rates of all the various possible dynamical pathways is beyond the scope of this paper and will be the subject of a future study. Note that the predicted gas production rates oscillate in time, suggesting a thermophysical mechanism for producing SW1's observed outbursts.

We quote and use the results of the sophisticated modeling from this point forward. The important connection between the sophisticated model treatment and the two simple limiting cases given previously is that the curves of "sophisticated model" $\tau_{AWI\ depletion}$ vs. $R_{nuc}$, trending approximately as $R_{nuc}^{1.63}$, fall between the small $\tau_{AWI\ depletion}$ ~ $R_{nuc}^1$ behavior predicted for energy delivery limited behavior and the large $\tau_{\text{AWI depletion}} \sim R_{\text{nuc}}{}^2$ behavior predicted for heat flow limited behavior. On the other hand, neither of the simple estimates accounts for potential internal sources of heat, and these can have a very large effect on the time until total AWI depletion, as can be seen by comparing the two curves spanning the literature range of possible heats of AWI -> CWI conversion in Fig. 1.

**2.2    Dynamical Timescales.** We now consider the movement of SW1 in the near-modern day solar system, where SW1 moved from its initial Kuiper Belt orbit to its present position in the Jupiter gateway region. The inward motion of a scattered disk KBO is initially started by self-stirring of the EKB, and not by planetary perturbations; but once started on a planet-crossing trajectory, a KBO can be scattered inward from one giant planet to the next on Myr timescales, with the largest amount of time spent furthest out by Neptune, where the orbital dynamical times are the slowest (Volk & Malhotra 2008, Bonsor & Wyatt 2012). While the exact dynamical path followed by an individual body depends sensitively on its initial state vector and the location of the planets (i.e., is highly chaotic), the median scattering time $\tau_{\text{dynamical}}$ is ~ 10 Myr (with 90% of objects in the range 5 – 20 Myr; Volk & Malhotra 2008, Prialnik & Rosenberg 2009, Sarid *et al.* 2019, Di Sisto & Rossignoli 2020, Gkotsinas *et al.* 2022).

This dynamical inward motion would have caused a concomitant large increase in $T_{\text{LTE}}$ for a KBO, increasing it from the 30-40 K typical of Kuiper Belt environments up to the 110-120 K found in the Jupiter Gateway region. Because dynamical and orbital timescales are much slower in the outer solar system (Kepler's laws), it takes ~70% of the dynamical migration time to move a typical object from the EKB at 30 – 60 au to ~10 au and $T_{\text{LTE}}$ ~ 90 K (Saturn's orbit). The rest of the time is spent by the body in the 90 – 110 K Jupiter-Saturn region, so we can expect processes happening at T < 90 K to be well completed during the ~10 Myr of dynamical migration, as well as any 90 – 110 K processes that take only a few Myr (like AWI conversion in km-sized "typical cometary" bodies, or the 4 small known "Gateway" objects; Table 1, Fig. 1; Fernandez *et al.* 2018). By contrast, 10 Myr is very short compared to any of the estimates (and especially the sophisticated model estimate of 60 – 100 Myr) for SW1's AWI conversion time. In fact, in 10 Myr SW1 is predicted to have transformed only 35 – 50% of its AWI by the sophisticated model (Fig 1a).

This implies that 29P/Schwassmann-Wachmann 1 is in a fundamentally different kind of internal state than a typical short period (SP) comet; i.e., it is large enough that the majority of its interior is not yet affected by its new thermal environment closer to the Sun. Whereas all potential AWI in comet nuclei of a more typical ~1 km size have likely fully crystallized prior to entering the Gateway or JFC populations, the sheer size of SW1 suggests that its migration into the Gateway would have provided insufficient time to fully crystallize all AWI present. Instead, SW1 may possess an actively advancing thermal front (~90 K according to the sophisticated Prialnik model), where the thermal environment of the Inner Solar System is slowly imprinting itself over the thermal environment of the Trans-Neptunian Population, and crystallizing AWI as it propagates. *This is more like the internal case of a Long Period (LP) comet, albeit with a rather thick layer of CWI near its surface.* This means that when SW1 transits the Gateway into the inner system in the next ~10 Kyr (Sarid *et al.* 2019), it will not only be the largest and brightest JF cometary object seen in the modern era, it could also be a highly unusual one.

### 3.  Comparison to Observations.
In the next four Sections we discuss how the proposed AWI -> CWI conversion model matches our current observational understanding of SW1. We start with a quick general summary of SW1's known behavior, then dive deeper into 4 different important individual issues.

The scenario of ongoing AWI crystallization driving SW1's activity is consistent with Wierzchos & Womack 2020's finding that the outbursts of dust and of CO from SW1 are not always correlated and that SW1 produces large amounts of gaseous CO into its coma but only negligible amounts of gaseous $H_2O$ (Ootsubo *et al.* 2012; Womack *et al.* 2017; Bockelee-Morvan *et al.* 2014, 2022). Unlike the situation for an active comet within ~ 2 au of the Sun (for which subsolar temperatures approach ~200K), at SW1's distance the bulk water ice matrix (+ dust + other ice impurities) that make up the comet nucleus are not very labile at the local equilibrium temperatures of ~110 K (Jewitt 2009; Lisse *et al.* 2021, 2022). Thus we expect neither large-scale bulk mass removal from a water ice dominated object, nor appreciable nuclear $H_2O$ gas emission (Figure 2). Instead, there should be removal of excess molecular "impurities" over and above the amount storable in a CWI hydrate lattice, $\lesseqgtr$ 20% of the total $H_2O$ ice volume (Schmitt *et al.* 1989, Jenniskens & Blake 1996), and a very low level of water gas production, both from the nucleus and days- to months-old icy coma dust that has had a long time to heat up in sunlight and sublimate as it flies away from the nucleus.

**4.    Dust Release.**  For SW1, we can distinguish 3 different types of emission activity from the nucleus: Dust + Gas (primarily CO) emission; Gas-only emission; and Dust-only emission. Each of these types is plausible for SW1 activity driven by AWI conversion, and a mix of all 3 behaviors could be what is creating the rather jumbled temporal history of SW1 quasi-periodic outbursts and quiescent behavior epochs observed over the last few decades.

Structural changes of bulk water ice structures due to minor species removal or AWI -> CWI volume phase changes can cause geomorphological rearrangements on icy bodies (c.f. the arguments made for Arrokoth's geomorphological structures by Moore *et al.* 2020 and Spencer *et al.* 2021). Such cave-in, landslide, and slumping sinkhole creation events could launch a small fraction of bulk material as "dust" (= rocky material + refractory ices & organics; Belton *et al.* 2008, 2011; Steckloff *et al.* 2016; Steckloff & Samarasinha 2018) outbursts as material fails and rearranges; as long as the rearrangement events create debris moving at velocities greater than $v_{escapeSW1} \sim 5.5$ m/s, that material will escape the surface and be launched into the coma[5]. Similar small-scale, localized mass-wasting processes have been proposed as capable of causing comet outbursts and maintaining activity (Steckloff *et al.* 2016; Steckloff & Samarasinha, 2018). As a result, SW1's frequent outbursts could thus be driven by frequent significant changes to its surface topography as trapped gas erupts and simultaneously blows off dusty surface material, a picture consistent with Ivanova *et al.* 2011's arguments that the surface layers of SW1 must be disrupted in order to create the amount of observed CO gas outflow.

Another possible mechanism for dust emission is CO gas entrainment of fine dust regolith. Unlike the case for JFCs in the innermost solar system, where even the bulk matrix water ice dominated material is subliming and releasing all its internal materials, including captured refractory dust, SW1's bulk water ice matrix is overall stable under the solid-solid phase AWI -> CWI crystallization process. Any dust will stay captured - until the previously mentioned shrinkage driven morphological rearrangements occur, and shards and pieces of water ice + rock are liberated. As long as these rearrangements produce sufficiently fine (submicron to micron-sized[6]) water ice + rock rubble, the copious amount of CO gas emission ( $\sim 4 \times 10^{28}$ mol/sec quiescent level or $\sim 1 \times 10^{29}$ mol/sec during outburst) can entrain this rubble and launch it into SW1's coma, especially on the day side near the sub-solar point where local temperatures are the highest (Fink *et al.* 2021). The two processes are coupled; liberation of fine dust and CO gas are both required for dust to be released by entrainment. If fine dust regolith production due to surface rearrangement is continual, then the dust production would appear to be governed by the production rate history of entraining gas (primarily CO), and CO + dust emission is observed. If instead the supply of loosely bound, fine regolith is produced stochastically and infrequently via episodic surface failures, then the rate of fine rubble production rather than the rate of CO production controls the observed dust outflow behavior, causing the two to appear decoupled. Sudden increase in nuclear CO production during an epoch where no fine dust regolith is available could produce CO-gas only outburst events. Conversely, if a sudden surface rearrangement (landslide, faulting, sinkhole creation, etc.) creates significant new reservoirs of fine dust after a long no-regolith epoch, then an apparent dust-only outburst could result during a CO-quiescent phase.

---

[5]Keane *et al.* 2022's detailed study shows that equatorial surface regions of a similarly sized object rotating with 16 hr period are gravo-rotationally unbound, so the required imparted velocity to cause dust escape could be less.

## 5.     Lack of Bulk CO Ice.

A counter argument to an object powered by AWI conversion that can be made is that the CO emission seen from SW1 is due instead to direct sublimation of large amounts of subterranean pure- or nearly pure CO ice. However, if CO-rich hypervolatile ice were abundant, it would sublime vigorously at $\sim 20K$ (less than its surface and interior temperature in the TNO population) and thus would have been lost during its billions of years of residency in the TNO population. Furthermore, the CO-rich ice should also be associated with $N_2$-rich phases, as $N_2$ gas is a species very similar in thermal sublimation properties to CO (Fray & Schmitt 2009, Lisse *et al.* 2021, Steckloff *et al.* 2021a) that was roughly abundant at the 10% level vs CO in the gas phase of the proto-solar nebula and in condensed phases of the PPD (Allamandola *et al.* 1992, Sandford & Allamadola 1993, Lacy *et al.* 1998. Kamata *et al.* 2019). Thus $N_2$ should be present in large quantities in SW1's coma if sublimation of large amounts of pure CO ice had survived intact from SW1's initial formation and were driving the observed CO production, but it is tellingly only seen at the ~0.01 level vs. CO (as deduced from $N_2^+$ measurements; Korsun *et al.* 2008, Ivanova *et al.* 2016, Womack *et al.* 2017).

---

[6]SW1's CO gas production rate, $Q_{CO}$, is ~4e28 molecules/s quiescent, up to $1 \times 10^{29}$ mol/s in outburst (Festou *et al.* 2001, Gunnarsson *et al.* 2008, Jewitt *et al.* 2008, Wierzchos & Womack 2020). $4 \times 10^{28}$ molecules/s is equivalent to a mass flux rate of $1.4 \times 10^{-7}$ kg/m²/s for a 32 km radius spherical body. Assuming ideal gas outflow at v = 400 m/s at $T_{LTE}$ = 115 K, this implies a local pressure of $8.6 \times 10^{-6}$ Pa above a 0.5 g/cm³ mean density body with surface gravitational acceleration = $4/3 G \rho R_{nuc}$ = $1.4 \times 10^{-3}$ m/s². Quiescent lofting of micron-sized particles of radius $a$ will occur when $F_{pressure} = \pi a^2 * 8.6 \times 10^{-6}$ Pa > $F_{grav}$ = $4/3 \pi a^3 \rho * 1.4 \times 10^{-3}$ m²/s or for $a < 3/4 * 8.6 \times 10^{-6}$ Pa/(500 kg/m³ * $1.4 \times 10^{-3}$ m²/s) = $9.2 \times 10^{-6}$ m = 9.2 µm, and up to $2.2 \times 10^{-5}$ m = 22 µm in outburst. For comparison, the typical JF comet at 1 au is capable of lofting up to ~10⁴ µm =1 cm sized particles. [We note also that simple, flat surface models of comet activity (in which gas drag must also overcome grain-grain surface cohesion) suggest that gas drag alone is incapable of lofting grains off the surface (*e.g., Gundlach et al. 2015; Jewitt et al. 2019*). Nevertheless, given that comets are known to actively emit dust, these models must not be complete; more modern models account for comet activity being tied to areas of steep topography (*Vincent et al. 2016*), and note that the grains may already be in motion due to e.g., mass wasting events (*Britt et al. 2004; Steckloff & Melosh, 2016; Steckloff & Samarasinha, 2018*); in this case, the gas does not need to overcome surface cohesion, but merely blow fine dust grains away.]

SW1 is thus very unlike hypervolatile rich comet C/2016 R2 (likely a rare case of an object preserving its original PPD composition of majority hypervolatile ices in the Oort Cloud; Lisse *et al.* 2021, 2022) with $N_2/CO \sim 0.1$ (Wierzchos *et al.* 2017, Biver *et al.* 2018, McKay *et al.* 2019), consistent with the predicted value of $N_2/CO \sim 0.06$ for icy planetesimals forming in the solar nebula at about 50 K (Owen & Bar-Nun 1995; Iro *et al.* 2003).

By contrast, AWI conversion would preferentially produce CO gas versus $N_2$ gas because CO's finite molecular dipole moment resulted in it becoming efficiently trapped in polar water ice phases Gyrs ago, forming a substantial CO molecule reservoir, while $N_2$'s zero homonuclear diatomic dipole moment means that it was very poorly retained by polar water ice phases making up the bulk of modern-day comets.

## 6. $CO_2$ Ice?

We have also considered bulk $CO_2$ ice as a possible source of the abundant CO gas emitted by 29P/SW1, but reject it for two reasons. While evidence for bulk $CO_2$ ice was found in the gas and ice emitted by the dying core of comet 103P Hartley 2, verifying models of $CO_2$'s survivability over 4.56 Gyr in the core of a small icy body (Davidsson 2021, Steckloff *et al.* 2021a, Lisse *et al.* 2022), there is little to no evidence of $CO_2$ emission from SW1 (Ootsubo *et al.* 2012, Harrington Pinto *et al.* 2022). This is despite the fact that $CO_2$ sublimes rapidly into vacuum at ~85K, about the same temperature that AWI ice recrystallizes (Fray & Schmitt 2009; Jewitt 2009; Lisse *et al.* 2021, 2022) - so we would expect both AWI and $CO_2$ to be mobilized in SW1. Nor is impurity phase CO stable in bulk $CO_2$ ice above 30K (i.e. the amorphous to crystalline transition for $CO_2$ ice occurs at 25-30 K; Escribano *et al.* 2013), making it only mildly more stable than the bulk CO removed within 1 - 40 Myr in KBOs (Davidsson 2021, Prialnik 2021, Steckloff *et al.* 2021a). The finding that JFC comets exhibit a strong correlation between their $H_2O$ and $CO_2$ (but not CO) production rates seen by Harrington-Pinto *et al.* (2022) is also highly consistent with $CO_2$ being housed in majority water ice phases.

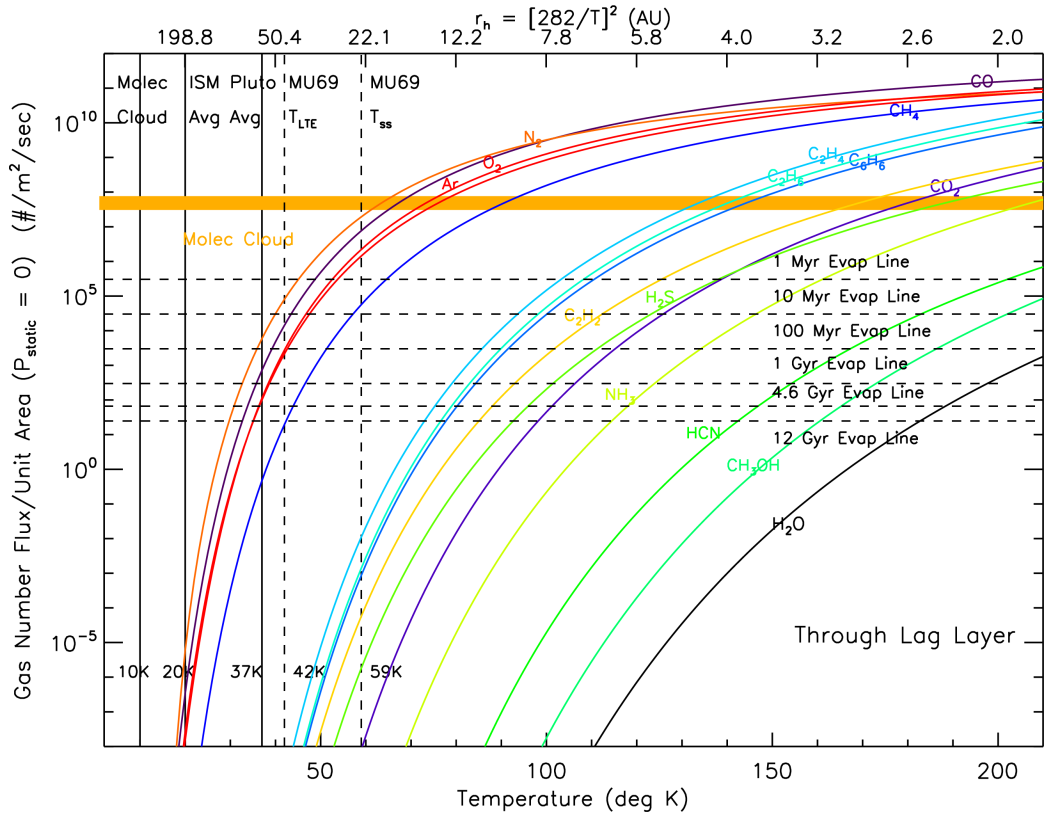

**Figure 3** – **Species specific $Q_{gas}$ vs Temperature curves for species expected in comets, Centaurs, and KBOs.** **Horizontal dashed lines**: values of the thermally driven outgassing rates at which an icy species is depleted in 1,10,100, 1000, 4600, and 12,000 Myrs for an Arrokoth-sized body. **Colored curves**: loss rates for a piece of ice of labeled composition evaporating at temperature T after allowing for an overlying lag layer with thermal diffusivity = $3x10^{-7}$ sec²/m impeding the flow of heat and gas into free space from the interior (Davidsson 2021, Steckloff *et al*. 2021, Prialnik 2021). **Top axis**: heliocentric distance from the Sun for a blackbody at local thermal equilibrium temperature T. From these curves and constraints, one can see that hypervolatile ices CO, $N_2$, and $CH_4$ are only stable in cold, dense molecular clouds and in modern KBOs residing beyond ~100 AU from the Sun, while metastable ices like $CO_2$ can survive in KBO and Centaur cores and hydrogen-bonded ices like $H_2O$ can even survive on inner system comet surfaces. After Lisse *et al.* (2022).

## 7. Lack of $CO_2$ Gas.

The almost total lack of $CO_2$ outgassing activity from SW1) *is* nevertheless surprising and may be a significant observational constraint. As stated above, if AWI is active, then we would expect $CO_2$ ice also to be active (Steckloff *et al.* 2015; Steckloff & Jacobson, 2016; Fig. 3). Further, there is good evidence for $CO_2$ outgassing activity at the 2-20% level vs. water at r < 2 au in multiple comets (Colangeli *et al.* 1999, A'Hearn *et al.* 2011, Bockelée-Morvan & Biver 2017, Läuter *et al.* 2019, Harrington-Pinto *et al.* 2022), suggesting that $CO_2$ should be abundant at the 2-20% level vs. $H_2O$ in *all* KBOs, Centaurs and comets[7].

However, the lack of $CO_2$ in SW1 may simply be due to its observed high CO abundance level. CO abundances can vary quite widely in comets, ranging from < 0.1 % up to ~15% vs $H_2O$ (as measured for comets at $r_h$ < 2.5 au when all 3 species are fully sublimating; Bockelée-Morvan & Biver 2017 and references therein; Harrington-Pinto *et al.* 2022). A'Hearn *et al.* 2012, noting an apparent anti-correlation between the abundance of CO and $CO_2$ in comets (especially in $CO_2$-ice and gas rich comet 103P/Hartley 2, the second target of the Deep Impact mission. A'Hearn *et al.* 2011), argued from surveying a large number of comets that the quantity [CO + $CO_2$] in comets appears to be conserved, and by implication, the CO and $CO_2$ found in KBOs, Centaurs and comets is sourced from the same reservoir. A recent survey of CO, $CO_2$ and $H_2O$ gas production in 25 comets by Harrington-Pinto *et al.* 2022 observed at $r_h$ < 2.5 au has confirmed this, finding a median value of [CO + $CO_2$] /[$H_2O$] = 18 +/- 4%. Whether this sourcing occurred in the proto-solar nebula/giant molecular cloud phase, or in the proto-planetary disk phase, is not yet clear; we do know that CO is highly abundant in the interstellar medium and, as the simpler molecule, was likely to have made up the majority of the original starting [CO+$CO_2$] reservoir (studies of dense molecular cloud core ices suggest [CO]/[$CO_2$]= 1 to 2, Suhasaria *et al.* 2017 & references therein).

On the other hand, the total amount of [CO+CO2] vs. water in short period comets, ~20% (Colangeli *et al.* 1999, A'Hearn *et al.* 2011, Bockelée-Morvan & Biver 2017, Läuter *et al.* 2019, Harrington-Pinto *et al.* 2022), is interestingly about the same as the total carrying capacity of crystalline water ice's pore space for minority impurities. By contrast, the study of dense molecular cloud core ices suggest that PPD ices had [CO + $CO_2$] /[$H_2O$] = 45 – 75 % (Suhasaria *et al.* 2017

---

[7] We note that 4.5 micron imaging data of SW1 obtained with Spitzer (Reach *et al.* 2013) and NEOWISE (Bauer *et al.* 2015) infrared space telescopes covered the combined emission bands of CO and CO2 in this regime. Unfortunately, the CO and $CO_2$ emission could not be separated and therefore no relative CO/$CO_2$ production rates could be obtained from these data. Although $CO_2$ production rates are cited in the tables of these papers, they are meant to be used as a proxy for the overall gas production rate and are not $CO_2$ production rates for SW1, a point explained in the original papers. The $Q_{CO2}$ values were derived assuming all gaseous emission in the images was due to $CO_2$, not allowing for a mix of the two volatiles. As stated in those papers, SW1 is well known to have substantial CO production rates and $Q_{CO2}$ can be converted to $Q_{CO}$ with a multiplicative factor of 11.6 (due to the difference in fluorescence efficiencies). Harrington Pinto *et al.* (2022) recently inferred $CO_2$ production rates from the Spitzer and NEOWISE imaging data using CO production rates from mm-wavelength spectra of CO obtained contemporaneously with the IRAM 30-m telescope. These inferred $CO_2$ production rates confirm that CO is produced in much higher quantity than $CO_2$ in SW1's coma.

& references therein). As the laboratory work of Moore *et al.* 1991 and Pilling *et al.* (2010) have showed that it is very easy to interconvert between CO and $CO_2$ embedded in $H_2O$ ice upon energetic charged particle radiation, this provides a plausible operative mechanism for producing large $CO_2$ abundances from initially high CO body abundances in icy bodies. If penetrating, ionizing radiation effects dominate, this could mean that the current $CO_2$ abundance in a comet is directly related to the ionizing radiation dose undergone by the original CO molecules in the original proto-solar ices and thereafter in the aggregated icy body (Pilling *et al.* 2022). Thus the relative $CO_2/CO$ production ratio inside ~8 au, when both CO and $CO_2$ sublime vigorously (Lisse *et al.* 2021, 2022; Fig. 3) could be a measure of the radiation exposure age of the $CO_2$-releasing ice, as well as the shielding effects of overlying layers[8].

The preponderance of CO emission coupled with a lack of $CO_2$ activity suggests another possible mechanism controlling the AWI conversion. It is known from laboratory studies and observations of ISM ices that the CO in water ice resides on the surface of pores and that these pores coalesce as temperature increases, keeping roughly the same total volume while reducing their surface area for gas adsorption (Jenniskens & Blake 1996, Palumbo 2005, Bossa *et al.* 2014, Cazaux *et al.* 2015, David *et al.* 2019, He *et al.* 2019). This pore surface loss is the mechanism whereby internal AWI can slowly lose CO as it warms up from the 30-40 K temperatures of a Kuiper belt to the ~90K temperatures where AWI -> CWI conversion occurs, and whereby large amounts of gas are evolved upon AWI -> CWI crystallization. What is not commonly understood in the literature is the difference in the behavior of adsorbed CO vs $CO_2$ molecules in AWI as it begins to crystallize. He *et al.* 2016 have shown that in the temperature range of 88 - 105 K, the sticking probability for CO to AWI is ~0, while that for $CO_2$ is in the range of 0.95-0.45. This will lead to preferential emission of CO and retention of $CO_2$ as AWI converts to CWI. What should thus be left behind as the thermal wave passes into the body are $CO_2$-rich water ice phases, with $CO_2$ comprising up to ~ 20% vs $H_2O$ in the thermally mature CWI material (i.e., up to the impurity species carrying capacity of CWI; this is consistent with Harrington Pinto *et al.* 2022's finding for 25 comets with r < 2.5 au that $CO_2$ outgassing may be strongly tied to water production). How well this differential sticking probability segregates the CO vs $CO_2$ will depend on how slowly the thermal wave warming up the AWI and converting it into CWI propagates into the body; as stated in Section 2, sophisticated models (Prialnik *et al.* 2004, 2008; Prialnik & Rosenberg 2009) have $\tau(R_{nuc}) \sim R^{1.63}$, $dR_{nuc}/dt \sim R^{-0.63}$, both for exothermic and non-exothermic AWI -> CWI conversion (Fig. 1), so larger bodies should segregate the two ices slower and more efficiently. Also consistently, the temperature of the phase front converting AWI to CWI in these numerical models is ~90K, close to the optimal temperature for differential adsorption separation of CO from $CO_2$ in AWI.

---

[8] Note that we are only discussing CO and $CO_2$ trapped in AWI here; the observational evidence does not support CO emitted from SW1 via conversion of free $CO_2$ ice into free CO ice, unless this is happening as an ongoing process today that is transforming the $CO_2$ ice to CO ice with near 100% efficiency (a highly unlikely process; Pilling *et al.* 2022), otherwise $CO_2$ gas would be detected in SW1's coma. This is because at the ~90K temperatures required to convert AWI to CWI (Jewitt 2009), $CO_2$ ice is also very volatile ($CO_2$ flash sublimes into vacuum at 85K, Escribano *et al.* 2013; the equilibrium saturation pressure is very large, Lisse *et al.* 2021; Fig. 3).

Proving which of these plausible mechanisms (or combination thereof) are operant in small icy solar system bodies – a high initial $CO/CO_2$ abundance ratio coupled with a total fixed amount of $CO + CO_2$ in comets; radiation driven conversion of CO <-> $CO_2$ ice in AWI; or solid state distillation via preferential sticking of $CO_2$ to AWI at 85 – 105 K -  should be possible using detailed studies of an SW1 that is still actively transforming its AWI nearby. This is precisely why the title of this paper suggests SW1 as an excellent laboratory for studying AWI -> CWI and CO <-> $CO_2$ conversion, and in the next two sections we elaborate on potential future detailed studies of conversion-related processes that could be performed.

## 8.    Remote Sensing Tests.

From the arguments made above, we find SW1's behavior makes a strong case for ongoing AWI crystallization, one that is very stochastic in its behavior, and one in which the structure of SW1 is constantly being rearranged as a thermal wave warming the interior from ~40  to ~120 K propagates through a highly heterogenous and/or fractally connected body. This conversion should continue for Myr timescales as the Gateway thermal wave propagates deeper into the interior [somewhere on the order of between (60 to 100 Myr thermal relaxation time) – (1 to 10 Myr dynamical emplacement time) = 50 to 100 Myr].

A number of remote telescopic investigations are immediately suggested by this line of reasoning. Further extension and testing of Jewitt 2009's and Li *et al.* 2020's Centaur activity vs heliocentric distance findings is of course warranted, in order to further refine our understanding of the energy of activation and any $\Delta H_{AWI->CWI}$ involved with SW1's activity. The total time for AWI conversion should scale roughly as the effective radius of the body to the 1.63 power (Prialnik *et al.* 2004, 2008; Prialnik & Merk 2008; and Prialnik & Rosenberg 2009; Section 2.1), so population surveys of Centaur activity in large vs small Centaurs with q < 10 au (e.g., like those of Fernandez *et al.* 2018 and Schambeau *et al.* 2021b) should be undertaken.

Also warranted are further searches for trends in CO versus $CO_2$ emission in Centaur and cometary bodies of known dynamical age. These would build on those carried out by Womack & Stern 1999, Wierzchos *et al.* 2017, and Harrington Pinto *et al.* 2022, but with emphasis on bodies of size between 6 and 30 km radius. If the CO to $CO_2$ conversion takes place inside the icy planetesimal (i.e., all icy solar system bodies started with reduced 100:0 ratios of $CO:CO_2$), then we can expect the largest bodies, like SW1, can shield their deep interiors & thus maintain their primordial CO molecules intact in their host $H_2O$ ice matrix, while the smallest bodies that lack much mass shielding, like $R_{nuc} \sim 0.7$ km hyperactive comet 103P/Hartley 2 and $R_{nuc} \sim 2.4$ km comet 67P/C-G, should have converted most of their CO into $CO_2$. If instead preferential adsorptive sticking of $CO_2$ to AWI at 88 – 105 K is the dominant operative mechanism, we can expect to find large amounts of $CO_2$- and water ice-rich dust in large Centaur comae that is very slowly evaporating (this could explain the small, but finite water gas emission and possible $CO_2$ gas emission seen for SW1 by Ootsubo *et al.* 2012; c.f. Womack *et al.* 2017 and arguments therein; Bockelee-Morvan *et al.* 2022). To be specific, future ground and space-based searches for water ice in SW1's coma and an extended coma icy dust source of water gas, ideally throughout epochs of both low and high emission activity, are warranted.

Parallel studies of Jupiter's Trojans as extremely old SW1 analogues are also warranted (e.g. Seligman *et al.* 2021), if they are indeed KBOs trapped into the L4 and L5 orbital resonances of the Sun-Jupiter system at r ~ 5.2 au at the time of the giant planet instability 3-5 Gyrs ago. In this case, while coming from the same small body feedstock, they would have Gyr ago converted all their AWI into CWI, and thus represent an AWI depleted endstate. In this way they could be as much akin to the JFCs as we argue that still ~50% AWI-constituent SW1 is like an Oort Cloud comet (Sec. 2.2).

## 9.    Future *In Situ* Spacecraft Measurements.
Our AWI -> CWI hypothesis also suggests tests to be done by *in situ* spacecraft directly investigating the chemistry and geomorphology of SW1. Previous experience with cometary in situ survey missions like Deep Impact and Rosetta have consistently turned up new regimes of behavior unobservable from the Earth. E.g., the sharp < 10 min rise times for comet 9P/Tempel 1's water sublimation rate upon night-time water ice frost rotating into sunlight (A'Hearn *et al.* 2005) or the neck striae and layering of 67P's lobes created by formation and evolutionary processes (Massironi *et al.* 2015, Ruzicka *et al.* 2019).

If we were observing the vigorous sublimation of ***bulk*** CO ice, we would expect it to occur with much surface mass removal and ejection of commingled dust, potentially producing the steep-walled pits seen on recent JFC comet 81P/Wild 2 (Brownlee *et al.* 2004) and KBO 2014 MU69 (Arrokoth; Stern *et al.* 2019, Singer *et al.* 2019a). The occasional extreme outburst could be due to pockets of CO gas buried beneath a lag layer requiring the buildup of significant over-pressure before being released in a stochastic outburst accompanying blow-out crater formation. Significant whole-body mass-wasting will produce a relatively young surface geomorphology.

CO released from AWI instead should be co-mingled with other minor impurities present, and their release could be searched for by an *in situ* gas analyzer [although if sourced from depth, these minor impurities would have to be able to percolate out through ($T_{SubSolar} + T_{midnight}$)/2 ~ (150+30)/2 = 90 K interior material (Huebner *et al.* 2006, Davidsson 2021, Lisse *et al.* 2021, Prialnik 2021, Steckloff *et al.* 2021a). Impurity release caused by an amorphous to crystalline water ice transition **will not** result in significant mass wasting, but subtler shrinkage effects due to the 2-3% molar volume increase between Low Density Amorphous (LDA) water ice and crystal Ih water ice (Fraser *et al.* 2004, Palumbo 2005, Tanakaa *et al.* 2019; this rises to an ~18% molar volume increase for transitions between High Density Amorphous (HDA) water ice and crystalline Ih water ice), producing relatively gentle "subsidence" geomorphologies like depressions and folds (but not craters). The material left behind by AWI conversion coupled with CO gas evolution should be high strength and rich in $CO_2$ + crystalline water ice, with a preserved (although somewhat relaxed) cratering impact record.

AWI conversion driven outbursts could occur in the same fashion as for bulk CO ice, as the breakthrough of buried overpressured pockets, or by surface failure creating collapses that expose fresh ice-rich material, as seen in JFC 67P (Vincent *et al.* 2015, Steckloff *et al.* 2016, Pajola *et al.* 2017). The time required to build up enough pressure to create an overburden-failure caused outburst, coupled with predicted periodic thermophysical gas production excursions (Fig. 2) could lead to quasi-periodic outgassing and outbursting behavior. This is in contrast to other outburst mechanisms such as those caused by landslides, which is a phenomenon primarily confined to the surface. Furthermore, the morphological manifestations of the outbursts caused by such different mechanisms would be different because of the associated geophysical differences. Therefore, nearly continuous high-resolution coma and surface morphology monitoring, ideally by an *in situ* spacecraft mission [like the proposed AMBITION (Bockelee-Morvan et al. 2021)], Centaurus (Singer *et al.* 2019b), or CHIMERA (Harris *et al.* 2019)], over extended timescales (e.g., months to years), should yield critical clues to help resolve the drivers of activity in this enigmatic object.

Measurement of coma dust particles, either by direct sampling or spectroscopic mapping, can also provide telltale information. The low level of water gas production in SW1's coma, only seen in outburst (Bockelee-Morvan *et al.* 2010, 2014, 2022), is likely due to the stability of crystalline water ice against sublimation at 6 au ($T_{LTE} \sim 115$ K, $T_{ss} \sim 161$ K; Ooutsubo *et al.* 2012, Lisse *et al.* 2021, Fig. 3). Therefore coma grains surrounding SW1 should still contain much of their original water ice unless they are very old (i.e., far from the nucleus). Whether this water ice will be contained in separate ice particles containing refractory ices like $H_2O$, HCN, and $CH_3OH$ (Lisse *et al.* 2021, 2022) or comingled as part of the "typical" dry cometary refractory ferromagnesian silicates and sulfides, future *in situ* sample collection and characterization measurements should find this "dust" to be very "wet" and water ice-rich. Finally, if the $CO_2$ preferential sticking mechanism of He *et al.* 2016 is important, this dust could also be $CO_2$-rich, at least in the inner coma regions where the dust has yet to warm up above $\sim 105$ K.

**10. Conclusions & Recommendations.**  In this paper we have investigated the nature of 29P/Schwasmann-Wachmann 1, a large and unusual cometary Centaur. We have utilized thermophysical models including conversion of water ice from an amorphous to a crystalline state to show how 29P could still be, in the current day, harboring a large fraction of the amorphous water ice it started out with in the Kuiper Belt, and that that amorphous water ice is currently undergoing crystallization within the deep interior, releasing highly volatile CO gas, and driving activity. We argue that this AWI -> CWI conversion from the lower density amorphous water ice to the higher density crystalline water ice would be accompanied mainly by a compressional volume decrease, but not a large mass wasting, causing the nucleus to shrink, inducing interior and surface structural failures, which could be driving the frequently seen outbursts of this comet. Using the relative abundances for the majority species CO, $CO_2$, and $H_2O$ seen in comets, we then show how 29P compares to other comets and fits into the evolutionary KBO -> Centaur -> Short Period comet evolutionary picture, concluding that 29P is evolving much slower than the "typical" km-sized centaur and will become a very unusual JFC, full of AWI, when it transitions through the Jupiter dynamical gateway (Sarid *et al.* 2019).

If correct, our arguments have a number of important, testable predictions, including: the quick release on Myr timescales of CO from AWI conversion for any few km-scale scattered disk KBO transiting into the inner system; that to date SW1 has only converted between 50 to 65% of its nuclear AWI to CWI; that volume changes upon AWI conversion could have caused subsidence and cave-ins, but not significant mass wasting on SW1; that SW1's coma should contain abundant amounts of CWI $CO_2$-rich "dust" particles; and that when SW1 transits into the inner system within the next ~1 Myr, it will be a very different kind of JF comet.

All of these findings are predicated on the assumption that AWI exists in KBOs and Centaurs, and that AWI conversion, like direct water ice sublimation, is a fundamental process that sculpts and alters icy bodies in the solar system. But unlike direct water ice sublimation that drives inner system comet activity, it is poorly studied. As the closest known natural example of a body actively undergoing AWI conversion, we thus strongly recommend further intense study of 29P/SW1. This study should utilize both remote sensing characterization and monitoring (c.f. Womack *et al.* 2020 and https://wirtanen.astro.umd.edu/29P/ 29P_obs.shtml) as well as direct exploration via a future *in situ* spacecraft mission [e.g., AMBITION (Bockelee-Morvan *et al.* 2021), Centaurus (Singer *et al.* 2019b), or CHIMERA (Harris *et al.* 2019)].

## 11. Acknowledgements. The authors would like to thank D. Jewitt, A. Poppe and S.A. Stern for many useful discussions that inspired and honed this work. C.M. Lisse was supported by the New Horizons mission project and J.K. Steckloff was supported by NASA grants 80NSSC18K0497 and 80NSSC19K1313 for the analyses reported in this manuscript. M. Womack and O. Harrington Pinto were supported by the National Science Foundation under grants AST-1615917 and AST-1945950, and this study is based in part on work done by M. Womack while serving at the National Science Foundation.